\documentclass[12pt]{article} 
\usepackage[utf8]{inputenc}
\usepackage{amsmath,amssymb,amsfonts} 
\usepackage[UKenglish]{babel} 
\usepackage[T1]{fontenc}
\usepackage{setspace}
\usepackage{xcolor}
\usepackage{graphicx}
\usepackage{lmodern}
\usepackage{booktabs}
\usepackage{ae,aecompl}
\usepackage{subfigure}
\usepackage[toc,page]{appendix}
\usepackage{epstopdf}
\usepackage{import} 

\usepackage{todonotes}
\usepackage{fullpage}
\allowdisplaybreaks 

\graphicspath{{figures/}}
\newcommand{\real}{I\!\!R}

\DeclareMathAlphabet{\mathpzc}{OT1}{pzc}{m}{it}

\def\be{\begin{equation}}
\def\ee{\end{equation}}
\def\l{\label}
\title{Wick Rotation in the Tangent Space} 
\author{Joseph Samuel$^{a}$\\ 
$^a$ Raman Research Institute, Bangalore 560 080}
\begin{document}

\maketitle

\begin{abstract}
Wick rotation is usually performed by rotating the time 
coordinate to imaginary values. In a general curved spacetime,
the notion of a time coordinate is ambiguous. 
We note here, that within the tetrad formalism of general relativity, 
it is possible to perform a Wick rotation directly
in the tangent space using considerably less structure: a timelike,
future pointing vector field, which need not be Killing or hypersurface
orthogonal. This method has the advantage of 
yielding {\it real} Euclidean metrics, even in spacetimes which are not static. 
When applied to a black hole exterior, the null generators
of the event horizon reduce to points in the Euclidean spacetime.
Requiring that the Wick rotated holonomy of the null generators be trivial 
ensures the absence of a `conical singularity' in the Euclidean space. To illustrate the basic idea,
we use the tangent space Wick rotation to compute the 
Hawking temperature by Euclidean methods in a few spacetimes including
the Kerr black hole. 

\end{abstract} 

\section{Introduction}
Euclidean spacetimes are analytic continuations of Lorentzian
ones. This complex rotation of the time coordinate to imaginary values
yields many insights into the quantum nature of space and time. 
Euclidean methods are motivated by the formal similarity 
between the propagator 
for quantum mechanics and the partition function in statistical mechanics. 
Results obtained by Euclidean methods agree with those obtained by completely
different techniqes, examples being the Hawking temperature and entropy  
of black holes\cite{PhysRevD.15.2752,Wald:1984rg}. See Ref. \cite{sorkin}
for an discussion of Euclidean methods and path integrals.

In the usual approach, we identify a time coordinate $t$ and 
perform the replacement
$t\rightarrow it$. In quantum field theory in flat space-time, the time coordinate of an
inertial observer is usually selected for Wick rotation. In a general curved spacetime, there need
be no preferred time coordinate. In stationary spacetimes, the Killing vector $\xi$ does give us a preferred notion
of a time {\it direction}, but does not in general give us a time coordinate. One can use a time function $t$
partly defined by the condition ${\cal L}_\xi t=1$. 
But, if $t$ is a time function, so is $t'=t+f(x)$, 
where $f(x)$ is any function constant along the integral curves of $\xi$. Coordinate 
Wick rotation in $t$ and $t'$ give us different metrics  which in general are complex.

In the special case of static spacetimes, 
the Killing vector $\xi$ is hypersurface orthogonal and
determines  a time function via $\xi_a=\lambda\nabla_a t$, where $t$ is determined up to affine transformations
$t\rightarrow at +b$, with $a,b$ constants. These transformations commute with Wick rotation and result in the same Euclidean
metric in transformed coordinates. This permits us to define coordinate Wick rotation (CWR) for static spacetimes. A good example
is the Schwarzschild exterior, where the timelike Killing vector is hypersurface orthogonal. The method results in a real Euclidean
metric and correctly gives the thermodynamic properties of the Schwarzschild black hole.
Except in special cases \cite{Wald:1984rg} 
like static spacetimes, the procedure
does not result in a {\it real} Euclidean metric.

The purpose of the present paper is to explore Wick rotation in the Tangent space, using
a tetrad frame. Wick rotation in flat space involves not only rotating the time coordinate,
but also correspondingly rotating tensor field components to imaginary values.  
In theories with spinor fields, Wick rotation  also entails rotating the spinor fields \cite{Waldron1998369,vanNieuwenhuizen199629}. 
As is well known, in order to handle spinors in curved spacetime, 
one has to introduce tetrads.   
One might as well formulate Wick rotation in the tangent space from the beginning using tetrads.
As we will see, our method leads to {\it real} Euclidean metrics. We do not need a globally defined time coordinate,
but can make do with slightly less structure. What we {\it do} need is a local notion
of ``time'' in each tangent space:
a timelike, future pointing vector field $u^i$. In the special case that $u^i$ is Killing 
($D_{(i}u_{j)}=0$) and hypersurface orthogonal ($u_i=\lambda D_i t$), 
these local notions of time mesh together to give a global time coordinate
and our method reduces to the usual Wick rotation. We shall also be
interested in black hole exterior solutions, where the timelike vector field becomes null on the boundary. 
Indeed this is the most interesting application of Euclidean methods. In this
case, we shall also require that the integral curves of the timelike
vector field lie in the exterior region.

In section II, we describe the formalism we use
and in section III, apply this to some simple two dimensional 
examples to illustrate the difference between coordinate Wick 
rotation (CWR) and tangent space Wick rotation (TSWR). 
We show that unlike CWR, TSWR yields real Euclidean metrics.
In section IV, we apply Tangent space Wick rotation to the 
Kerr black hole, arrive at a real Euclidean metric  and compute 
the Hawking temperature. Section V is a summary.

\section{Wick Rotation in Frames}
Let $(g,\cal{M})$ be a Lorentzian spacetime. 
The real metric $g_{ij}$ {i,j=0,1,2,3} can be locally 
expressed in real orthonormal tetrads $e^a,{a=0,1,2,3}$
\be
g_{ij}=e^a_ie^b_j \eta_{ab},
\l{tetrad1} 
\ee
where $\eta_{ab}$ is the Minkowski metric.
The tetrad fields uniquely define the $SO(3,1)$ connection 1-form $A^a{}_b$
via the Cartan structure equation:
\be
d e^a+A^a{}_b\wedge e^b=0
\l{spinconnection}
\ee
Suppose now that we are also given a timelike future pointing vector field
$u^i$. Let us choose an orthonormal tetrad frame so that $e^0=\hat{u}$, 
which is normalised to $\hat{u}.\hat{u}=-1$. (Our metric signature 
is mostly plus) .  
A Wick rotation in the tangent space consists of the replacement
$e^0\rightarrow ie^0$, which of course results in a real Euclidean metric.
\be
{\mathcal G}_{ij}=e^a{}_ie^b{}_j \eta_{ab}=e^0{}_ie^0{}_j+e^1{}_ie^1{}_j+e^2{}_ie^2{}_j+e^3{}_ie^3{}_j
\l{tetrad2} 
\ee
and the Cartan Structure equation yields a Euclidean $SO(4)$ 
connection ${\mathcal A}{}^a{}_b$, which differs from the Lorentzian connection 
only in that the space-time components in the internal indices
are multiplied by $-i$: ${\mathcal A}
{}^a{}_b =A^a{}_b$ if $a,b=1,2,3$ and 
${\mathcal A}
{}^a{}_b =-i A^a{}_b$ if $a$ or $b$ is $0$. ($A^0{}_0$ vanishes since the connection preserves
the Minkowski tensor $\eta_{ab}$). Thus the holonomy of the Wick rotated 
frame is just the Wick rotation of the holonomy of the Lorentzian connection.
The relation between the Euclidean and Lorentzian metrics can also be expressed 
without reference to frames:
\be
{\mathcal G}_{ij}=g_{ij}+2{\hat u}_i{\hat u}_j,
\l{ger}
\ee
where the reality of ${\mathcal G}$ is manifest.

An interesting application of Wick rotation is a black hole exterior ${\cal E}$,
the region from which escape to infinity is possible. This region has a null
boundary ${\cal N}=\partial {\cal E}$, the event horizon, which has signature $(0,+,+)$. The null boundary
${\cal N}$ is ruled by null generators. We suppose the black hole 
exterior is axisymmetric as obtains for the Kerr metric and indeed 
all stationary black holes.
Since we require that $u^i$ is a timelike vector field whose integral curves remain in ${\cal E}$, it
follows from $u.u<0$ that $u$ must approach a null generator of ${\cal N}$ on the boundary.

Let us choose a $u^i$ timelike
everywhere,  but degenerating to null at ${\cal N}$. We can also choose it to
agree with the coordinate time translation at infinity.
Performing a Wick rotation 
leads to a Euclidean metric in the bulk. But at the horizon, we find that each
null generator $N$  of ${\cal N}$ projects down to a single 
point $p(N)$ in the Euclidean 
space . A timelike curve $C$ that just grazes the horizon close to $N$ must 
therefore correspond to a closed curve encircling
this point. This demands a periodic identification of the ends of $C$
in the Euclidean space and thus an identification of the whole space by
an isometry. 
In order to avoid a conical singularity at
the image of the $N$, we must demand that the Wick rotated holonomy of the 
null generator be trivial. 
Indeed, that the holonomy is $H_N=\exp{2\pi J}=1$,
where $J$ is a generator in the Lie Algebra of $SO(4)$. 
As we will see this condition will give us the Hawking
temperature of the black hole.

\section{Some Two Dimensional Examples}
In this section we describe two simple toy  examples to bring 
out the difference between CWR and TSWR. 
These are all essentially two dimensional examples, 
in which the calculations and visualisation are easy. 
Of course we must replace $SO(3,1)$ and $SO(4)$
above by $SO(1,1)$ and $SO(2)$ to make the correspondence with the two dimensional case.
These are all cases
in which both methods CWR and TSWR are possible, (because in two dimensions, stationary
spacetimes are also static). But to make our point we use non 
static Painleve-Gullstrand coordinates. 
\subsection{Rindler and Painleve-Gullstrand}
Consider Rindler spacetime (with $g$ an arbitrary positive constant, $R>0$)
\be
ds^2=-g^2R^2dT^2+dR^2
\l{rindler}
\ee The CWR replacement give us 
\be
ds^2=g^2R^2dT^2+dR^2
\l{eucrindler}
\ee
The horizon at $R=0$ is a null curve that goes over to a point in the Euclidean spacetime.
Requiring the absence of a conical deficit at $R=0$ fixes $T$ to be
a periodic coordinate with period $\beta=2\pi/g$. This gives the correct 
Unruh temperature
$T_U=g/(2\pi)$.

Consider now  the metric 
\be
ds^2=-dt^2+(dr-v(r)dt)^2
\l{pgr}
\ee
where $v(r)=\sqrt{1-2gr}$ and the range of coordinates is $-\infty<t<\infty,0<r<1/(2g)$. 
(One can cover Rindler spacetime with two coordinate patches, the Painleve form for $r< 1/(2g)$ 
and the Rindler form
overlapping with it and continuing to  the region beyond.)
In these coordinates, the  metric appears stationary and 
not static, because of the presence of the cross terms $dr dt$. 
In fact, the metric (\ref{pgr}) is simply a part of Rindler spacetime (\ref{rindler}) expressed in Painleve-Gullstrand
coordinates, which are non singular on the horizon, $r\rightarrow 0$. 
The coordinate transformation 
$t=T+f(R),r=gR^2/2$, where 
\be
f(R)=-1/g[\sqrt{1-g^2R^2}+\log{gR}-\log({1+\sqrt{1-g^2R^2}})] 
\ee
relates the two metrics
(\ref{pgr}) and (\ref{rindler}).
If we were to use CWR and replace $t$ with $it$ in (\ref{pgr}),
we would get a complex metric from the terms linear in $dt$. 

However using TSWR, we set $e^0=u=dt$, $e^1=(dr-v(r)dt)$ and the Euclidean metric 
has the real form
\be
ds^2=dt^2+(dr-v(r)dt)^2
\l{epgr}
\ee

\begin{figure}
\begin{center}
\begin{tabular}[c]{cc}
  \begin{tabular}[c]{c}
\includegraphics[scale=0.5]{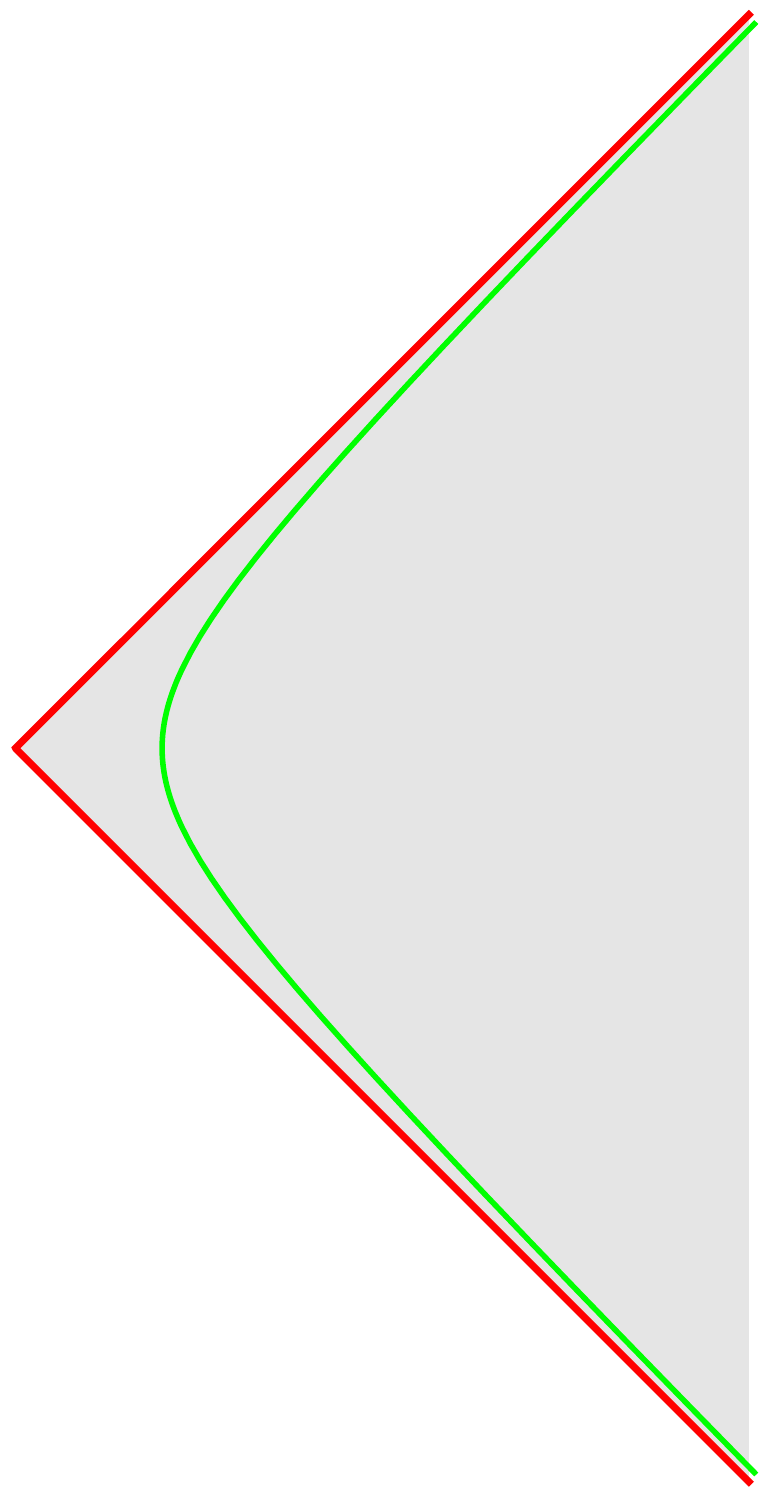}        
  \end{tabular} &
  \begin{tabular}[c]{c}
\includegraphics[scale=0.5]{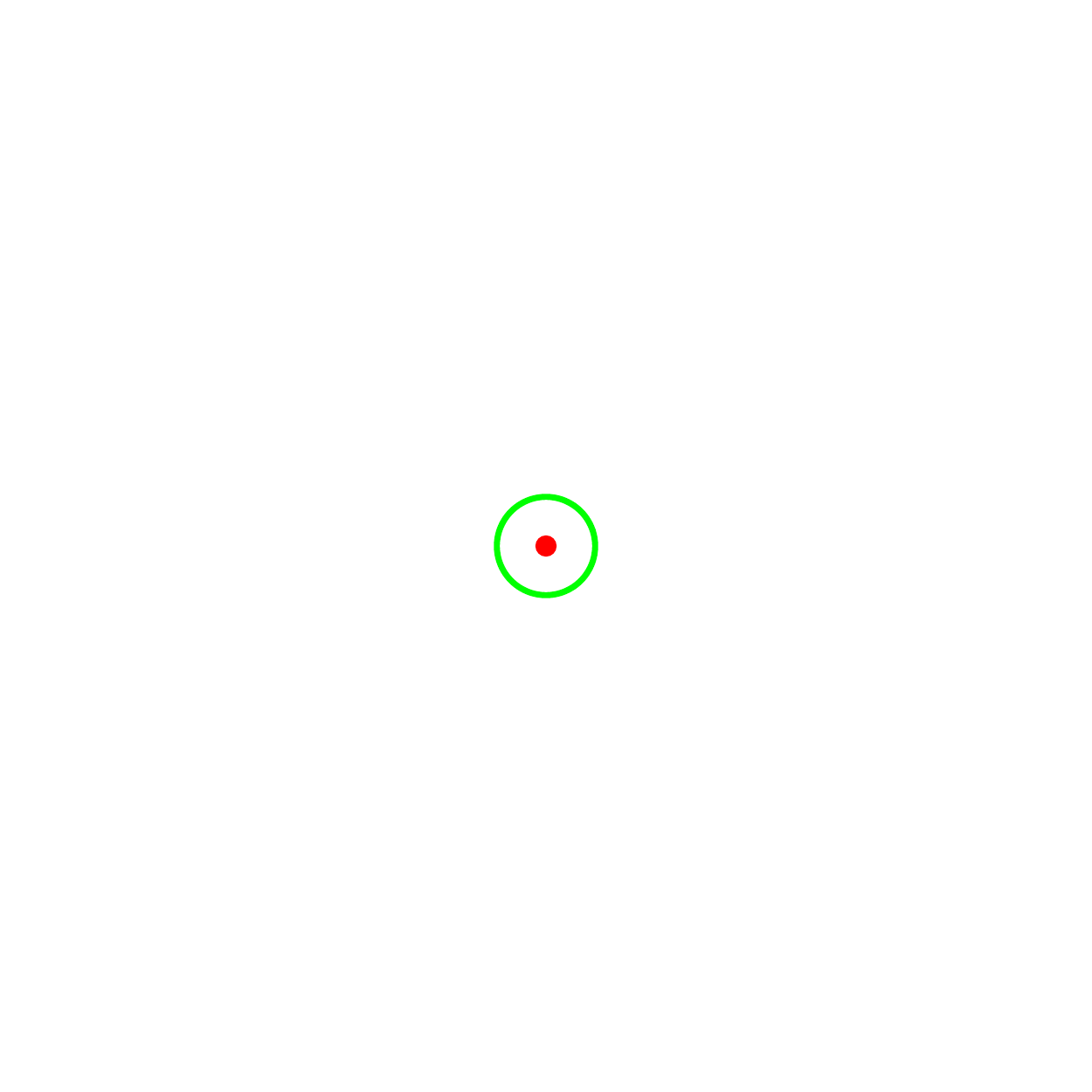}
  \end{tabular} \\
  (a) & (b) \\
\end{tabular}
\end{center}
\caption{Fig. 1a shows part of the right Rindler wedge, shaded in gray. 
The red curve is the null generator of the Rindler horizon. 
The green curve is a timelike
curve that just grazes the horizon. Figure 1b shows the Euclidean transcription of Fig. 1a. The null
generator of the horizon has collapsed to a point (shown in red) and the green
timelike curve close to it, must have its ends identified to remain close to
the point in red. This forces us to make periodic identifications in the time direction. }
\end{figure}

The line at $r=0$ is a null generator of the horizon
and goes over to a point in the Euclidean spacetime. This also removes
the ``kink'' in the horizon at the origin of Minkowski space.
From (\ref{spinconnection}) we work out the Lorentzian $SO(1,1)$ connection as 
\be
A=-v'(r)(dr-v(r)dt) K,
\label{Aexpression}
\ee 
where $K$ is the boost, the $2\times2$ Pauli matrix $\sigma_1$, which
generates the group.
Computing the holonomy of the 
Lorentzian connection along
the null curve $r=0$, we find 
\be
H(A)=\exp{\int_0^\beta K (v'(r))(-v(r)dt)}=\exp[-(v^2(r))'\beta K/2]
\l{holonomy}
\ee
where $\beta$ is the range of the integration.
We require that the holonomy of this null curve when continued to Euclidean values be trivial. i.e. 
we impose the condition 
\be
-(v^2(r))' \beta/2=2\pi
\label{vsqeq}
\ee
or  $\beta^{-1}=T_U=g/(2\pi)$ as before.

\subsection{Schwarzschild and Painleve-Gullstrand}
Very similar considerations apply to the Schwarzschild metric. By spherical symmetry, this
is essentially a two dimensional situation
\be
ds^2=-(1-2M/r)dt^2+(1-2M/r)^{-1} dr^2 +r^2d\Omega^2
\l{sch}
\ee
CWR results in a real Euclidean metric\cite{Wald:1984rg} and the correct Hawking 
temperature $T_H=1/(8\pi M)$.

However, if we present the Schwarzschild metric in Painleve-Gullstrand form
\be
ds^2=-dt^2+(dr-v(r)dt)^2+ r^2 d\Omega^2
\l{pgs}
\ee
with $v(r)=\sqrt{2M/r}$
the CWR would result in a complex metric. But if we set
$e^0=u=dt,e^1=(dr-v(r)dt)$ and then compute the Euclidean metric,
it takes the form
\be
ds^2=dt^2+(dr-v(r)dt)^2+ r^2 d\Omega^2
\l{epgs}
\ee
The computation of the Holonomy is then idential to the earlier case. We find that 
(\ref{Aexpression})  is still true, but now with $(v(r))^2=2M/r$.
The condition that the Euclideanized  holonomy of the null generator is trivial 
gives us again (\ref{vsqeq}). Using the new form of $v^2$ 
gives us the correct Hawking temperature. In both the above examples, it was possible
to transform to static coordinates and use CWR rather than TSWR. The spacetimes were static,
but viewed in non static coordinates.

In both these examples, we had a Lorentzian manifold which was geodesically 
incomplete. Geodesic completion of the Lorenztian manifold would 
result in recovering the full spacetime - all of Minkowski 
space in the Rindler example (3.1) 
and  the maximally extended\footnote{The curvature singularity 
obstructs a smooth completion.}  Schwarzschild spacetime in (3.2).
However if we perform a Wick rotation, we get a Euclidean metric which 
is also geodesically incomplete. Completion then leads to a smooth
Riemannian manifold, {\it provided certain global identifications are made}.
These identifications involve modding out by an element of the isometry group
and leads to a periodic time coordinate and a well defined temperature.
The Euclidean completion of the Schwarzschild exterior 
has no counterpart of the Schwarzschild interior or the curvature singularity.

\section{The Kerr Black Hole}
We now deal with the Kerr metric which is genuinely non static. In fact, the situation
is slightly worse: the metric is not even globally stationary, since the Killing vector which
is timelike at infinity turns spacelike in the ergoregion. We apply the same procedure,
by computing the Holonomy along the null generators of the Kerr horizon and requiring
that its Euclidean continuation be trivial. The computation of the holonomy is
entirely straightforward if a little tedious. For clarity, we supress the details and present only the
starting point and the final expressions. 

The calculation is aided by 
explicit formulae given in Chandrasekhar\cite{chandrabook}, 
whose notation we use below. The useful short forms
$\Delta(r)=r^2+a^2-2Mr=0$, $\bar{\rho}=r+ia\cos{\theta}$ and 
$\rho^2=r^2+a^2\cos{\theta}^2$ are standard in the Kerr metric. 
Note however that our metric signature is opposite to Ref.\cite{chandrabook}. 
The Kerr metric in standard Boyer-Lindquist coordinates is described by 
the null tetrads
\begin{eqnarray}
l=1/\Delta (r^2+a^2,\Delta,0,a)\\
n=1/\rho^2(r^2+a^2,-\Delta,0,a)\\
m=1/(\bar{\rho}\sqrt{2}) (ia\sin{\theta},0,1,i{\rm cosec}{\theta})\\
\l{kerrtetrads}
\end{eqnarray}
from which we find the inverse metric
\be
g^{ij}=(-l^i n^j -n^i l^j+m^i\bar{m}^2j+\bar{m}^i m^j)/2
\l{kerr}
\ee

The null generators of the event horizon at $\Delta(r)=0$ 
are tangential to $n$. 
We transform from null tetrads to Minkowskian ones by the transformation
$e^0=(l+n)/\sqrt{2},e^3=(l-n)/\sqrt{2},e^1=(m+\bar{m})/\sqrt{2},e^2=(m-\bar{m})/(\sqrt{2}i)$.
Writing $n^i=\frac{dx^i}{d\tau}$, where $\tau$ is a parameter along the null generator, 
we need the integral
\be
\int_0^\tau n^i A_i{}^a{}_b=\tau n^iA_i{}^a{}_b,
\label{integral}
\ee
along a stretch of the null generators of the Kerr horizon,  where $A^a{}_b$ is defined by
(\ref{spinconnection}) .
This matrix $F^a{}_b=n^iA_i{}^a{}_b$ is a  
$4 \times 4$ matrix 
\[ \left( \begin{array}{cccc}
0 & \mu & \nu_1&-\nu_2 \\
\mu & 0 & \nu_1&-\nu_2 \\
\nu_1 & -\nu_1 & 0&0\\
-\nu_2 & \nu_2 & 0&0 
\end{array} \right)\] 
where $\nu=\nu_1+i\nu_2=\frac{ia \sin{\theta}}{\sqrt{2} (\bar{\rho})^2}$ and 
$\mu=(r-M)/\rho^2$. 
The eigenvalues of this matrix are $\{\mu,-\mu,0,0\}$ corresponding to eigenvectors $v_1,v_2,v_3,v_4$.
$F^a{}_b$ generates the transformation
\be
F=\mu K,
\ee
where $K$ is a boost $\sigma_1$ in the $v_1-v_2$ plane. The  $v_3-v_4$ plane is left invariant by this
transformation.

From the condition of trivial Euclidean holonomy we find
\be
\mu \tau=2\pi
\ee
We can express the parameter $\tau$ in terms of $\beta$ the $t$ difference between the endpoints of C

\be
\int_0^\tau n^t d\tau=\beta=\frac{r^2+a^2}{\rho^2}\tau
\ee
Eliminating $\tau$ in favour of $\beta$ gives 
\be
\beta^{-1}=T_H=\frac{r-M}{2\pi(r^2+a^2)} 
\ee
where all quantities are evaluated at the image of the horizon.
This expression gives the correct Hawking temperature.

\begin{figure}[h!t]
\includegraphics[scale=0.5]{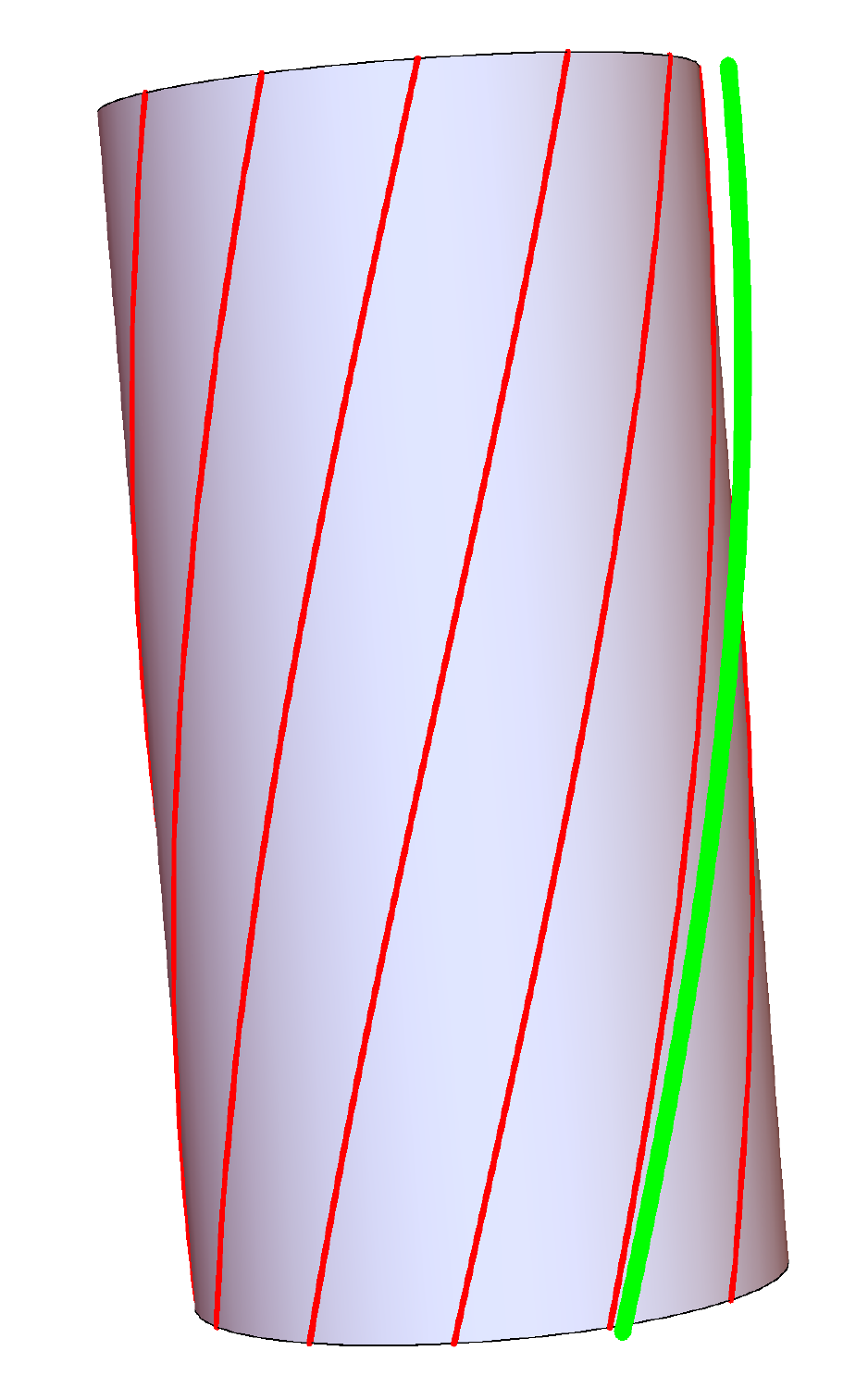}
\caption{Figure shows the Kerr horizon with the $\theta$ coordinate supressed for visualisation. Some generators
of the horizon are shown in red and a timelike curve just outside the horizon is shown as a thick green line. In the Euclidean spacetime,
this curve must be compactified to a circle.}
\end{figure}

Geometrically, the Kerr exterior ${\cal E}$ is a geodesically incomplete
spacetime, since there are infalling geodesics that leave ${\cal E}$ within
a finite affine parameter (or proper time). Completing ${\cal E}$ would 
lead to the maximally extended Kerr spacetime \cite{chandrabook}. 
If we Euclideanise the Kerr exterior, along the lines above, we find again that
the Euclidean metric is also geodesically incomplete. The completion of 
the Euclidean metric gives us a smooth Euclidean manifold provided
certain global identifications are made. All the points along a null
generator of the horizon have to be identified as a single point in the 
Euclidean manifold. Thus the Euclidean image of the $S^2\times \real$
Kerr horizon is just an $S^2$, which has codimension 2. Geodesics which
meet this $S^2$ normally define a planar geometry very similar to the $t-r$
plane of Schwarzschild. The requirement that the holonomy of a small circle
encircling this point is trivial gives us the Hawking temperature.
\section{Conclusion}
Whether or not one wants to consider complex metrics is largely 
a matter of taste. Especially in dealing with the Kerr metric and the Newman-Penrose formalism, 
considerable use is made of complex analytic techniques. While these techniques have their value,
the physical interpretation often demands that we deal with real metrics. It is entirely possible to use 
CWR to analyse the Kerr metric, by going into the complex domain. 
Our main point here is that a real domain analysis
is also possible.

In the Lorentzian Kerr geometry, the horizon has topology $S^2\times\real$, where $\real$ represents the null generators. In the 
Euclidean version the image of the horizon is an $S^2$ embedded in four dimensional Euclidean space. Since $S^2$ has codimension two,
the picture is locally as shown in Fig. 1b, where the $S^2$ is represented by a point. 
Timelike curves in the Lorentzian geometry that graze the horizon map to curves encircling the horizon. 
A timelike three surface  with topology $S^2\times \real$  
just outside the horizon ($\Delta$ slightly positive) maps to a three manifold of topology $S^2\times S^1$, which encircles the image of the Lorentzian horizon.
As would be expected, the region within the 
Lorentzian horizon disappears entirely in the Euclidean description.

One sometimes writes CWR as $t\rightarrow (\exp{i\theta}) t$, where $\theta$ varies from $0$ to $\pi/2$ and interpolates
between Lorentzian and Euclidean manifolds. In the case of TSWR we would have 
$e_0\rightarrow (\exp{i\theta}) e_0$. The main difference is that CWR complexifies the manifold, while TSWR
complexifies the (tetrad) fields living on the manifold. If $e_0$ is hypersurface orthogonal ($e_{0a}=\lambda \nabla_a t$), one may try to view this
as complexifying t, the time coordinate. However, this interpretation works only if $\frac{\partial}{\partial t}$ is Killing. Otherwise the metric 
would also have $t$ dependence and we would need to complexify $t$ in the metric functions as well. TSWR works even when $u$ is neither Killing
nor hypersurface orthogonal over the manifold.  

One may wonder about the role of $u$ and the arbitrariness involved
in the choice of $u$. $u$ provides us with a local notion of `time', 
which in semiclassical gravity determines a division of modes into positive
and negative frequencies. (See for instance Ref. \cite{deutschcandelas} 
which deals with field theory near a boundary.) 
Note however, that it is only the behaviour
of $u$ in the neighbourhood of the black hole horizon that enters
into our discussion. In this neighbourhood, $u$ is essentially equal to
the vector field whose integral curves are those of a locally non rotating
observer. The arbitrariness in the choice of $u$ away from the horizon does
not affect the calculation of the Hawking temperature.

Another situation in which the CWR fails is the Black hole interior. 
This region is not static even for the Schwarzschild black hole.
However, the geometrical ideas of this paper still apply. Let us supress the angular coordinates for simplicity.
The region has a singularity at the origin and a null generator of the horizon at $r=2M$.
The null generator becomes a point in the Euclidean spacetime.
A {\it spacelike} curve just {\it inside} the horizon must therefore encircle this point.
The requirement of trivial holonomy yields the same value
for $\beta$ as in the exterior. In this case the entire exterior region disappears from the Euclidean description. 
We have a Euclidean disc, whose centre is the image of the horizon and the singularity runs around the rim of the disc.
The corresponding picture for the Kerr case is far from clear. Even for the Schwarzschild interior, the physical
interpretation of this Euclidean picture is unclear.

We have shown that Wick Rotation does not need a global 
notion of time. A timelike vector field is enough. Such a vector field induces a local notion of time in each tangent space.
In static situations
these local notions of time mesh together to create a global time coordinate,
but there are interesting situations where this does not happen.  
In black hole exteriors, where there is a null boundary,
the idea of Wick rotation in the tangent space can be used to derive 
real Euclidean metrics.  The condition that the Euclideanised holonomy
of the null generators of the boundary be trivial ensures that there is no
conical singularity in the Euclidean space and correctly 
gives the Hawking (or Unruh) temperature 
in the examples we have considered.

\section{Acknowledgements}
It is a pleasure to thank Abhay Ashtekar, Vishwambhar Pati, T.R. Ramadas  
and Sandipan Sengupta for discussions.



\bibliographystyle{utphys}

\providecommand{\href}[2]{#2}\begingroup\raggedright\endgroup

\end{document}